\documentclass[onecolumn,journal,twoside]{IEEEtran}
\usepackage[utf8]{inputenc}

\usepackage[T1]{fontenc}
\usepackage{url}
\usepackage{ifthen}
\usepackage{cite}

\usepackage{graphicx}
\usepackage{amssymb}

\usepackage[cmex10]{amsmath} % Use the [cmex10] option to ensure complicance
\usepackage{amsthm}
\usepackage{cite}

\newtheorem{theorem}{Theorem}

\theoremstyle{remark}
\newtheorem{remark}{Remark}

\usepackage{xcolor}
\usepackage{subcaption}

 \newcommand{\bF}{\mathbb{F}}

\title{Entangled Polynomial Codes for Secure, Private, and Batch Distributed Matrix Multiplication: Breaking the ``Cubic'' Barrier}
		\author{Qian~Yu, and A.~Salman~Avestimehr\\
		 Department of Electrical and Computer Engineering\\ University of Southern California, Los Angeles, CA, USA
		 \thanks{This work has been submitted to the IEEE for possible publication. Copyright may be transferred without notice, after which this version may no longer be accessible.}}

\begin{document}

\maketitle

\begin{abstract}
%In distributed matrix multiplication, an important scenario 
In distributed matrix multiplication, a common scenario is to assign each worker a fraction of the multiplication task, by partitioning the input matrices into smaller submatrices. In particular, by dividing two input matrices into $m$-by-$p$ and $p$-by-$n$ subblocks, a single multiplication task can be viewed as computing linear combinations of $pmn$ submatrix products, which can be assigned to $pmn$ workers. Such block-partitioning based designs have been widely studied under the topics of secure, private, and batch computation, where the state of the arts all require computing at least ``cubic'' ($pmn$) number of submatrix multiplications. Entangled polynomial codes, first presented for straggler mitigation, provides a powerful method for breaking the cubic barrier. It achieves a subcubic recovery threshold, meaning that the final product can be recovered from \emph{any} subset of multiplication results with a size order-wise smaller than $pmn$.   In this work, we show that entangled polynomial codes can be further extended to also include these three important settings, and provide a unified framework that order-wise reduces the total computational costs upon the state of the arts by achieving subcubic recovery thresholds.  
\end{abstract}

\section{Introduction}

%challenges of straggler, security, and privacy, and emerging field of coded computing

%1 or 2 pages explain the scheme

Large scale distributed computing faces several modern challenges, in particular, to provide resiliency against stragglers, robustness against computing errors, security against Byzantine and eavesdropping adversaries, privacy of sensitive information, and to efficiently handle repetitive computation \cite{dean2013tail, 10.1145/357369.357371, 10.1145/357172.357176, 5663489, 7958569, pmlr-v89-yu19b, so2019codedprivateml}. Coded computing is an emerging field that resolves these issues by introducing and developing new coding theoretic concepts, started focusing on straggler mitigation \cite{lee2015speeding,dutta2016short,NIPS2017_7027}, then later extended to secure and private computation  \cite{chang2018capacity, 8382305,  pmlr-v89-yu19b, kim2019private,chang2019upload, so2019codedprivateml}.

Coding for straggler mitigation is first studied in \cite{lee2015speeding} for linear computation, where classical linear codes can be directly applied to achieve same performances. For computation beyond linear functions, %the error correcting codes no longer commute with the computation, and 
new classes of coding designs are needed to achieve optimality.
%To mitigate the straggler effect, where some of the workers may delay or fail to complete their assigned computation. This is first studied in \cite{lee2015speeding} for linear computation, where classical linear codes can be directly applied to achieve the same performances. For computation beyond linear functions, the error correcting codes no longer commute with the computation, and new classes of coding designs are needed to achieve the optimal performances.
%Stra started by linear comp, generalized to mat mat [poly], and poly comp [lcc]. Then PCC. Poly has been used for pairwise...
%(combination para 2,3)
%One of the earliest works in coded computing is . One basic scenario is to mitigate the  straggler effect, where some of the workers may delay or fail to complete their assigned computation. This is first studied in \cite{lee2015speeding} for linear computation, where classical linear codes can be directly applied to achieve the same performances. For computation beyond linear functions, the error correcting codes no longer commute with the computation, and new classes of coding designs are needed to achieve the optimal performances.
In \cite{NIPS2017_7027}, we studied matrix-by-matrix multiplication and introduced the polynomial coded computing  (PCC) framework. The main idea is to jointly encode the input variables into single variate polynomials where the coefficients are functions of the inputs. By assigning each worker evaluations of these polynomials as coded variables, they  
% and assign After computation is applied onto these coded variables, each worker 
essentially evaluate a new polynomial composed by each worker's computation and the encoding functions at the same point.   As long as the needed final results can be recovered from the coefficients of the composed polynomial, the master can decode the final output when sufficiently many workers complete their computation. %.receiving results from. %Explicitly, whenever the master receives computation results from any subset of workers with a size greater than the degree of the composed polynomial, 
%the composed 
%the final output can be decoded by interpolating the composed polynomial from its evaluations. %, and then proceed to decode the final output.
%The property is referred to as achieving a recovery threshold of the same quantity.
PCC significantly reduces the design problem of coded computation to finding polynomials %with a composed polynomial
satisfying the above decodability constraint while minimizing the degree of the decomposed polynomial. It has been shown that PCC achieves a great success in providing exact optimal coding constructions for large classes of computation tasks including: pair-wise product \cite{NIPS2017_7027}, convolution \cite{NIPS2017_7027, 8437563}, inner product \cite{8262882, 8437563}, element-wise product \cite{8437563}, and general batch multivariate-polynomial evaluation \cite{pmlr-v89-yu19b}. 

An important problem in distributed matrix multiplication is to consider the case where the inputs are % distributedly compute matrix-matrix multiplication, where the inputs are
encoded and multiplied in a block-wise manner. This setup generalizes the problem formulated in \cite{NIPS2017_7027} to enable a more flexible tradeoff between resources such as storage, computation and communication, and has been studied in \cite{8437563, 8262882,  DBLP:journals/corr/abs-1901-07705, aliasgari2019private, 8613446, jia2019cross, jia2019generalized}. For straggler mitigation, the state of the art is achieved by two versions of the entangled polynomial code, both first presented in \cite{8437563}, which characterizes the optimum recovery threshold within a factor of $2$. For brevity, we refer the collection of them as \emph{entangled polynomial codes}.      
One significance of entangled polynomial codes is that it maps non-straggler-mitigating linear coded computing schemes to bilinear-complexity decompositions, which bridges the areas of fast matrix multiplication %algorithms
and coded computation, enables utilizing techniques developed in the rich literature (e.g., \cite{Strassen1969, 4567976,doi:10.1137/0120004,laderman1976noncommutative,WINOGRAD1971381,Bini1980,Sch_nhage_1981,doi:10.1137/0211020,Coppersmith:1981:ACM:1398510.1382702,Strassen:1986:AST:1382439.1382931,COPPERSMITH1990251, landsberg2006border, stothers2010complexity,Drevet2011,Williams12multiplyingmatrices,burgisser2013algebraic, Smirnov2013, %blaeser2018a,
sedoglavic2017non,sedoglavic:hal-01572046,lecomplexity}).  
Moreover, this connection reduces block-wise matrix multiplication to computing element-wise products, for which we developed the optimal strategy for straggler mitigation. % the augmented computation task is reduced 
The coding gain achieve by entangled polynomial codes extended to fault-tolerant computing \cite{8949560} %The gain for fault tolerance and security against Byzantine adversaries can be achieved by extending the concept of hamming distance to coded computing \cite{8949560}.
 %by generalizing the concept of hamming distance to coded computing,
 and it is shown in \cite{pmlr-v89-yu19b} that security against Byzantine adversaries can also be provided the same way. %, and xx is the same

The goal of this paper is to extend entangled polynomial codes to three main problems: secure, private, and batch distributed matrix multiplication. In secure distributed matrix multiplication \cite{chang2018capacity, 8382305, DBLP:journals/corr/abs-1810-13006, DBLP:journals/corr/abs-1812-09962, 8613446, DBLP:journals/corr/abs-1901-07705, kim2019private, 8675905 ,chang2019upload, 8761275, nodehi2019secure, jia2019capacity, kakar2019uplinkdownlink, aliasgari2019private, d2019degree}, the goal is to compute a single matrix product while preserving the privacy of input matrices against eavesdropping adversaries; in private distributed matrix multiplication \cite{kim2019private,chang2019upload, aliasgari2019private, 8832193}, the goal is to multiply a single pair from two lists of matrices while keeping the request (indices) private; batch distributed matrix multiplication \cite{jia2019capacity, jia2019cross, jia2019generalized} considers a scenario where more than one pair of matrices are to be multiplied.

There are recent works on each of these problems that considered general block-wise partitioning of input matrices \cite{DBLP:journals/corr/abs-1901-07705, aliasgari2019private, 8613446, jia2019cross, jia2019generalized}. However, all results presented in prior works are limited by a ``cubic'' barrier. Explicitly, when the input matrices to be multiplied are partitioned into $m$-by-$p$ and $p$-by-$n$ subblocks, all state of the arts require the workers computing at least $pmn$ products of coded submatrices per each multiplication task. 

%(TODO) what is subcubic?

We demonstrate how entangled polynomial codes can be extended to break the cubic barrier in all three problems. We show that the coding ideas of entangled polynomial codes and PCC can be applied to provide unified solutions %to also include %achieving order-wise improvements for handling many challenges including fault tolerance, 
%providing security and privacy to the computation, and efficiently handling batch 
with needed security and privacy, as well as efficiently handling batch evaluation.
%  In this work, we focus on demonstrating the gain in secure (against eavesdropping workers), private, and batch distributed matrix multiplication.
Moreover, we achieve order-wise improvements upon state-of-the-art designs with explicit coding constructions.
%Solving the basic straggler mitigation problem, 
%Entangled poly

%The two class of coding constructions, collectively refereed to as entangled polynomial codes, first presented in [], order-wisely outperforms related works on matrix multiplication with general block partitioning.  

\section{Overview of Coded Computation and Entangled Polynomial Codes}\label{sec:pre}

%there is a dataset x1...xk

A main challenge in distributed computing is to design schemes to operate in the presence of stragglers, which are workers that are slow or fail to return their computing results. Commonly, stragglers are handled using ``uncoded repetition'', where the same computation tasks are duplicated and assigned onto multiple worker machines. 
%We are interested to design a scheme that can tolerate stragglers. Common approach is repetiton. coded is recent approach to blah, explain, key idea. is to compute on coded data. maybe they provide an effictive way to blah.
Coded computation has recently been proposed as an effective approach to mitigate the straggler effect, and computing strategies has been proposed for a variety of computation tasks, including matrix multiplication \cite{lee2015speeding, dutta2016short, NIPS2017_7027, 8262882, 8437563}, convolution \cite{8006960, NIPS2017_7027}, Fourier transformation \cite{8262778, jeong2018coded}, element-wise multiplication \cite{8437563}, and %general batch
multivariate-polynomial evaluation \cite{pmlr-v89-yu19b}. The main idea of coded computing is to %instead
%design coding-theoretic frameworks to
assign each worker data or tasks in carefully designed coded forms, such that the final result can still be recovered after possibly non-linear computation is applied on coded data.
%and the main challenge is to ensure that 

%blah. compute on coded data, to provide effective way cite[]. variety of computation.

%It has been shown that by assigning each worker variables in coded forms, one can achieve significantly improved straggler resiliency compared to conventional uncoded replication \cite{lee2015speeding}. However, the key challenge is to design coding functions when $f$ and $g$ are nonlinear, to still ensure %the workers' results still allows
%efficient recovery of final results after nonlinear computation is applied on possibly coded variables.

%studied %in distributed computing frameworks
%in the context of straggler mitigation, where an unknown subset of workers could fail to return their computing results.

\begin{figure}[htbp]
 %	\vspace{-3mm}
\centering
  \includegraphics[width=0.7\linewidth]{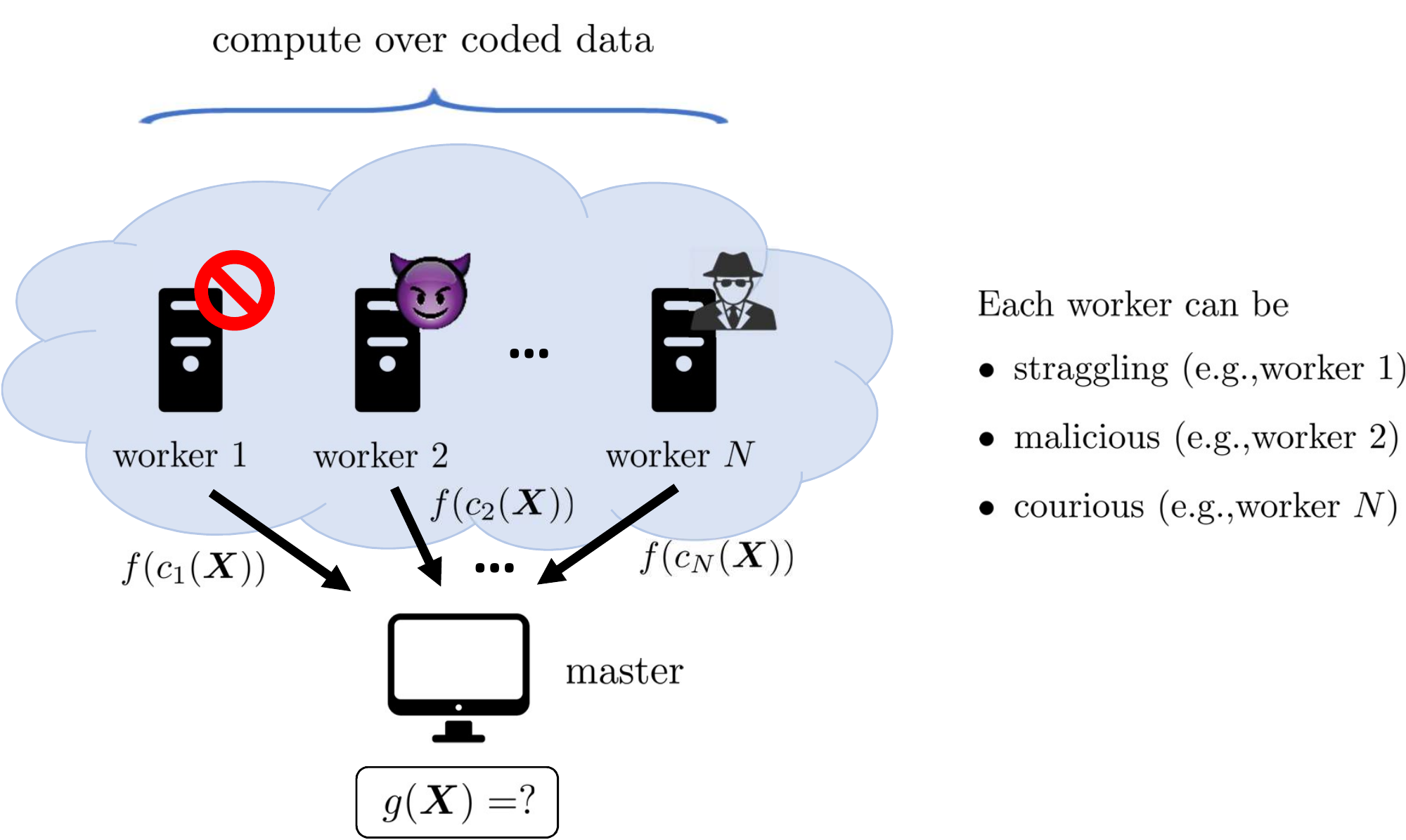}
\caption{An illustration of coded computation. A collection of workers aim to compute a function $g$ given an input dataset, where each worker can return an evaluation of a function $f$ with possibly coded data assignments. By carefully designing the coding functions ($c_i$'s), the final results can be efficiently recovered after computation is applies on coded data, in the presence of stragglers, while %system requirements including
providing security and privacy against malicious and colluding workers. \label{fig:overview}}
\label{fig:cent_compare}
\end{figure}

In a standard framework of coded computation (illustrated in Figure \ref{fig:overview}), we aim to design a computing scheme to compute a function $g$ over an input dataset $\boldsymbol{X}$ using $N$ distributed workers. Each worker computes a single evaluation of some function $f$, which can be viewed as building blocks of computing $g$. A conventional approach in distributed computation is to assign each worker an \emph{uncoded} fraction of the input dataset, and to recover the final results from evaluations of these uncoded assignments. %As shown in Figure \ref{figa}, in an uncoded computation design, the input dataset $\boldsymbol{X}$ is partitioned into $K$ (possibly overlapping) subsets, denoted $X_1$,...,$X_K$. Each worker $i$ computes $f(X_i)$, and $g(\boldsymbol{X})$ is to be decoded given $f(X_1)$,...,$f(X_K)$.   
We present two examples as follows:

\textbf{Matrix multiplication (column-wise partition)\cite{NIPS2017_7027}.} Consider a scenario where the goal is to compute the product $A^\intercal B$ given two large matrices $A\in \bF^{s\times t}$ and $B\in \bF^{s\times r}$. Here the input dataset is $X=(A,B)$, and the computation task is $g(A,B)=A^\intercal B$. After partitioning the input matrices column-wise into $m$ and $n$ submatrices of equal sizes, denoted $A_0,...,A_{m-1}$ and $B_{0},...,B_{n-1}$, the final results are essentially the collection of all $mn$ pair-wise submatrix-products $A_{i}^{\intercal}B_{j}$'s. If each worker can compute a single submatrix product of same sizes, i.e., $f$ is the multiplication of 
%Each worker can compute the product of two inputs with the same sizes of these submatrices, i.e., multiply 
two matrices of sizes $\bF^{\frac{t}{m}\times {s}}$ and $\bF^{{s}\times \frac{r}{n}}$, an uncoded design using $K=mn$ workers can be constructed by assigning the workers distinct $(A_i,B_j)$'s as inputs.  

\textbf{Polynomial evaluation\cite{pmlr-v89-yu19b}.}  Another example is to evaluate  multivariate polynomials on a dataset $X=(X_1,...,X_K)$. Explicitly, given a general polynomial $f$, the goal is to compute $g(\boldsymbol{X})=(f(X_1),...,f(X_K))$. If each worker can compute a single evaluation, then an uncoded design using $K$ workers can be obtained by assigning each worker $i$ input $X_i$.

%For example for block-partitioned matrix multiplication, the goal is to distributedly multiply matrices with sizes of $\bF^{t\times s}$ and $\bF^{s\times r}$, for a sufficiently large field $\bF$, %with a set of $N$ workers each 
%and each worker can multiply a pair of possibly coded matrices with sizes  $\bF^{\frac{t}{m}\times \frac{s}{p}}$ and $\bF^{\frac{s}{p}\times \frac{r}{n}}$. By partitioning the input matrices into $m$-by-$p$ and $p$-by-$n$ subblocks, a product of two input matrices can thus be viewed as linear combinations of $pmn$ submatrix products according to block-matrix-multiplication rules, which can be computed using $pmn$ workers with uncoded inputs.

%one can achieve significantly improved straggler resiliency compared to conventional uncoded replication \cite{lee2015speeding}. However, the key challenge is to design coding functions when $f$ and $g$ are nonlinear, to still ensure %the workers' results still allows
%efficient recovery of final results after nonlinear computation is applied on possibly coded variables.

A coded computing design that uses $N$ workers first encodes the dataset $\boldsymbol{X}$ %is first encoded 
using $N$ encoding functions $c_1,...,c_N$, then assign $c_i(\boldsymbol{X})$ to each worker $i$ as the coded input. In the presence of stragglers, the decoder waits for a subset of fastest worker until $g(\boldsymbol{X})$ can be recovered given the returned results from the workers. 
We say a coded computing scheme achieves a \emph{recovery threshold} of $R$, if the master can correctly decode the final output given the computing results from \emph{any} subset of $R$ workers. %\footnote
%{Note that recovery threshold is an equivalent measure of straggler resiliency, because it is identical to the difference between the number of workers and the number of stragglers that can be tolerated.}
This is an equivalent measure of number of stragglers that can be tolerated. 
The goal is to design the encoding functions\footnote{To ensure the complexities of the encoding and decoding functions are reasonably low, most related works focuses on linear coding designs.} to achieve the minimum possible recovery threshold given $f$, $g$, and $N$.

%We also want to ensure that the complexities of the encoding and decoding functions are reasonable low, 

\subsection{Polynomial Coded Computing (PCC)}

%However, 
A key challenge in coded computing is to design coding functions when $f$ and $g$ are nonlinear, to still ensure %the workers' results still allows
efficient recovery of final results after nonlinear computation is applied on coded variables. The PCC framework was introduced in \cite{NIPS2017_7027}, which achieves optimal recovery thresholds %system requirements
for large classes of functions including matrix multiplication, convolution, and polynomial evaluation \cite{NIPS2017_7027,8437563, pmlr-v89-yu19b}. 
%by encoding the input variables using carefully designed polynomials.
A general PCC design encodes the input dataset by assigning the workers evaluations of a carefully designed single variate polynomial. More specifically, the coding design is 
based on the following design parameters:
\begin{itemize}
\item A single variate polynomial $c(\cdot)$, where the coefficients are possibly random functions of the input variables.
\item $N$ evaluation points denoted $y_1,...,y_N$ from the base field. 
\end{itemize}
Then each worker $i$ obtains $c_i(\boldsymbol{X})=c(y_i)$ as the encoded variable. 
%Each worker $i$ selects a disjoint evaluation point in the base field, denoted $y_i$, and obtains an encoded variable that equals $c(y_i)$. 

After the workers apply $f$ on their assignments, they essentially evaluates the composed polynomial $f(c)$ at the same point. Hence, if the evaluation points  $y_1,...,y_N$  are distinct, and the decoder receives results from sufficiently many (%any subset with 
at least the degree of $f(c)$ plus one) workers, they can recover all information about polynomial $f(c)$. Based on this observation, the design problem in PCC framework is to construct a polynomial $c$, satisfying\footnote{
As well as other possible requirements such as complexity constraints on encoding and decoding functions (e.g., linear codes)\cite{NIPS2017_7027,8437563}, and data-privacy \cite{pmlr-v89-yu19b}.}
\begin{itemize}
    \item Decodability: the final result can be computed using coefficients of $f(c)$, 
\end{itemize} 
while minimizing the degree of $f(c)$ to achieve minimum recovery thresholds.
%optimizing for the following goal:
%minimizing the 
   % \item Maximum resiliency: 
 %  the degree of $f(c)$ to achieve maximum straggler resiliency.
   %is reduced as far as possible, . 
 %  \begin{itemize}
 %   \item Goal: 
%\end{itemize}
We present the following illustrative examples to demonstrate how PCC is applied to construct optimal codes for two example scenarios described earlier in this section.

\subsubsection{Polynomial codes \cite{NIPS2017_7027} for matrix multiplication}

%\textbf{Polynomial codes \cite{NIPS2017_7027}} for matrix multiplication: 
In the first example, the goal is to multiply two column-wise partitioned matrices $A=[A_0,...,A_{m-1}]$ and $B=[B_{0},...,B_{n-1}]$. To compute $A^\intercal B$, we essentially need to collect all $mn$ pair-wise submatrix-products $A_{i}^{\intercal}B_{j}$'s. 
%To recover all pairwise products $A_i^\intercal B_j$'s, 
Polynomial codes encodes the input dataset using the following polynomial: $c(x)=(\sum_{j=0}^{m-1}  A _j x^{j},\sum_{j'=0}^{n-1}  B _{j'} x^{j'm})$. More explicitly, given any distinct evaluation points $y_1,...,y_N$, each worker $i$ obtains  $(\tilde{A}_i,\tilde{B}_i)$ where
	\begin{align}
		\tilde{A}_i=\sum_{j=0}^{m-1}  A _j y_i^{j},\ \ \ \ \ \ \ \ 
	    \tilde{B}_i=\sum_{j'=0}^{n-1}  B _{j'} y_i^{j'm}  .
	\end{align}
   
   	\begin{figure}[htbp]
			\centering
			\includegraphics[width=0.8\linewidth]{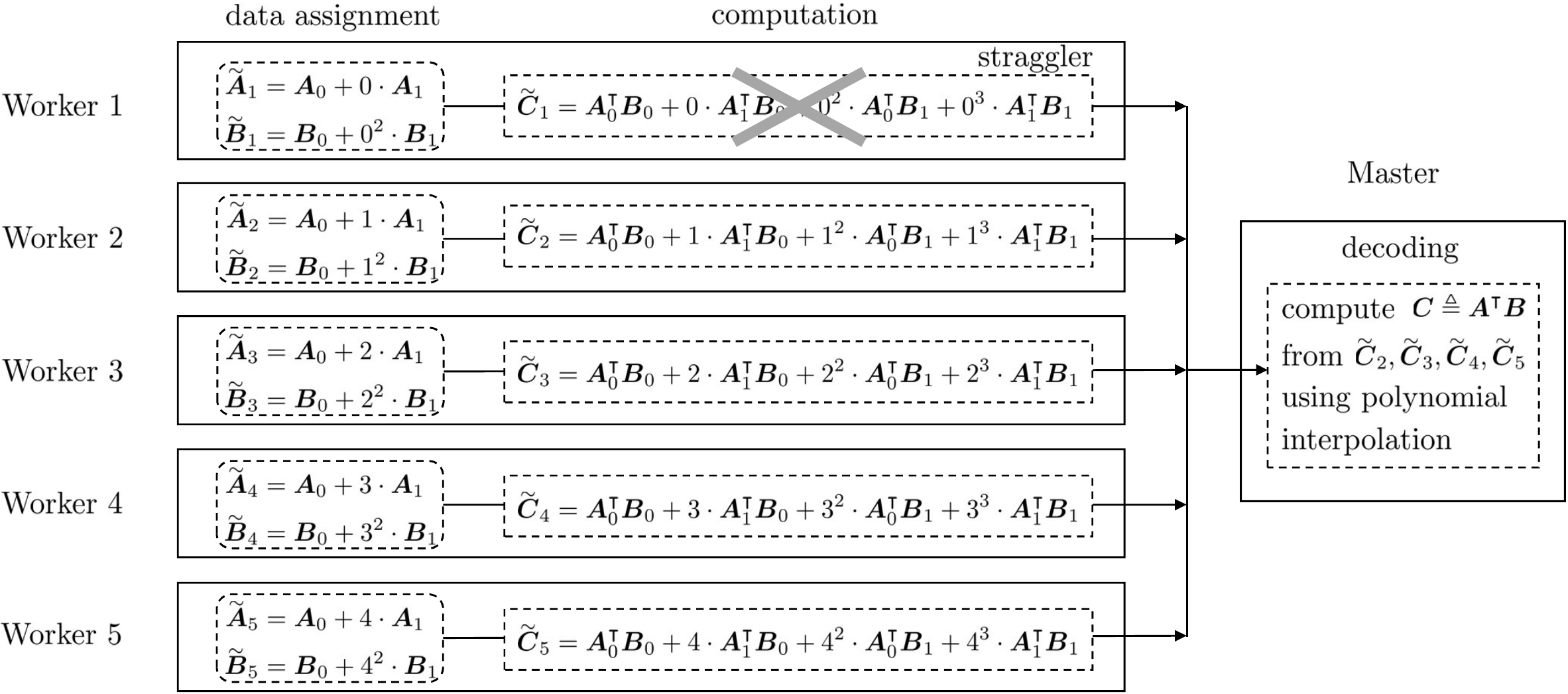}
			\caption{An illustration of polynomial code for matrix multiplication using $5$ workers that can each store half of each input matrix. The final result can be decoded from \emph{any} $4$ workers. 
			}
			%\vspace{-3mm}
			\label{p_poly:fig:exp}
		\end{figure}   
   
After the workers multiplies the assigned coded matrices, they essentially evaluates the composed polynomial $f(c(x))=\sum_{j=0}^{m-1}\sum_{j'=0}^{n-1}  A _j^\intercal  B _{j'} x^{j+j'm} $, which has degree of $mn-1$ and the $mn$ coefficients are exactly the $mn$ needed submatrix-products. After computation results are received from \emph{any} subset of $mn$ workers, the final results can be recovered from polynomial interpolation. I.e., achieving a recovery threshold of $mn$. It is proved in \cite{NIPS2017_7027} that polynomial codes achieves the optimal recovery threshold for this scenario. Polynomial codes was later extended to allow for general block-wise partitioning of the input matrices, as discussed in Section \ref{subsec:ent}.   
%poly is colwise, later generlized in [][] to allow for

%This property was also referred to as achieving a \emph{recovery threshold} of $mn$, which is an equivalent measure of number of stragglers that can be tolerated.

\subsubsection{Lagrange Coded Computing (LCC) \cite{NIPS2017_7027} for polynomial evaluation}

%   \textbf{Lagrange Coded Computing (LCC) \cite{NIPS2017_7027}} for polynomial evaluation: 
In the second example, the goal is to evaluate a polynomial $f$, of which the total degree is denoted $\textup{deg} f$. In particular, given an input dataset $X=(X_1,...,X_K)$, we aim to compute $f(X_1),...,f(X_K)$. 
   For straggler mitigation, LCC encodes the input variables using the  Lagrange polynomial $c(x)\triangleq \sum_{j\in[K]}X_j\cdot \prod_{k\in [K]\setminus\{j\}}\frac{x-x_k}{x_j-x_k}$ where $x_1,...,x_K$ are some arbitrary distinct elements from the base field $\bF$. In other words, each worker $i$ obtains the following $\tilde{X}_i$ as the coded variable.
   \begin{align}
    \tilde{X}_i\triangleq \sum_{j\in[K]}X_j\cdot \prod_{k\in [K]\setminus\{j\}}\frac{y_i-x_k}{x_j-x_k}.
\end{align}

   	\begin{figure}[htbp]
			\centering
			\includegraphics[width=0.9\linewidth]{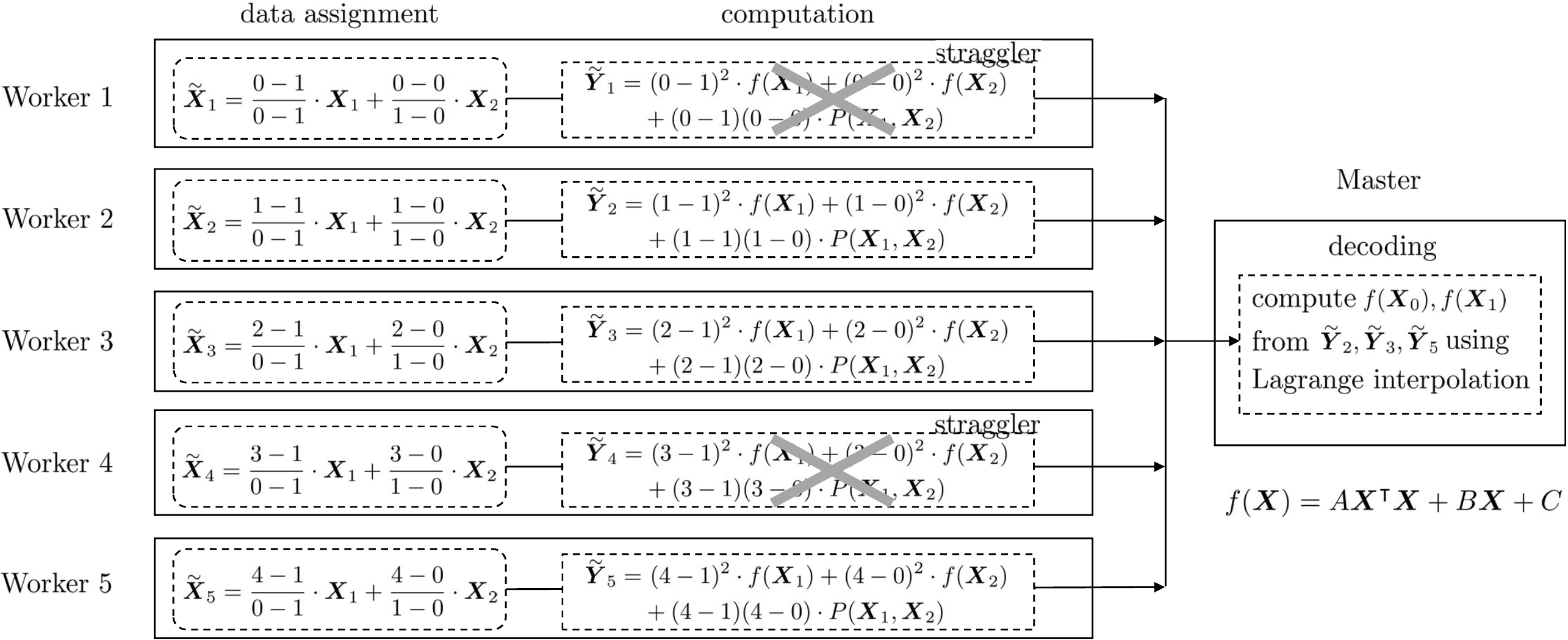}
			\caption{An illustration of LCC for evaluating a general degree-$2$ multivariate polynomial $f(X)=AX^\intercal X+BX+C$ using $5$ workers with an input dataset $\boldsymbol{X}=(X_1,X_2)$   that consists of two square matrices. The final result can be decoded from \emph{any} $3$ workers. 
			}
			%\vspace{-3mm}
			\label{lcc:fig_exp}
		\end{figure} 

After each worker applies function $f$ over the coded inputs, they essentially evaluates the composed polynomial $f(c)$, of which the evaluations at $x_1$,...,$x_{K}$ are exactly the $K$ needed final results. Let $\textup{deg} f$ denotes the total degree of polynomial $f$, the degree of the composed polynomial equals $(K-1)\textup{deg} f$. By assigning each worker distinct evaluation points, the decoder can recover all final results by receiving results from any subset of   $(K-1)\textup{deg} f+1$. This exactly achieves the optimum recovery thresholds among all linear codes when the number of workers $N$ is sufficiently large, while in other cases the optimum recovery thresholds are achieved by an uncoded version of LCC, where the evaluation points $y_1,...,y_N$ are selected from $x_1,...,x_K$ \cite{pmlr-v89-yu19b}. It was shown in \cite{pmlr-v89-yu19b} that by padding the input dataset with random keys, LCC simultaneously provides security against malicious workers and privacy of data against colluding workers, and achieves the optimal tradeoff between straggler resiliency, security, and privacy.

~

   PCC provides several other properties of interests: linearly-coded constructions of polynomial $c$ leads to linear codes; the encoding/decoding costs due to polynomial evaluation and interpolation can be handled using efficient algorithms with almost linear complexities \cite{von2013modern}.

%PCC is also for security and privacy

\subsection{Entangled Polynomial Codes}
\label{subsec:ent}

%todo
%
%Polynomial code was proposed for column-wise partitioning one can look at more general partitioning

\begin{figure}[htbp]
			%\vspace{-3mm}
			\centering
			\includegraphics[width=0.95\linewidth]{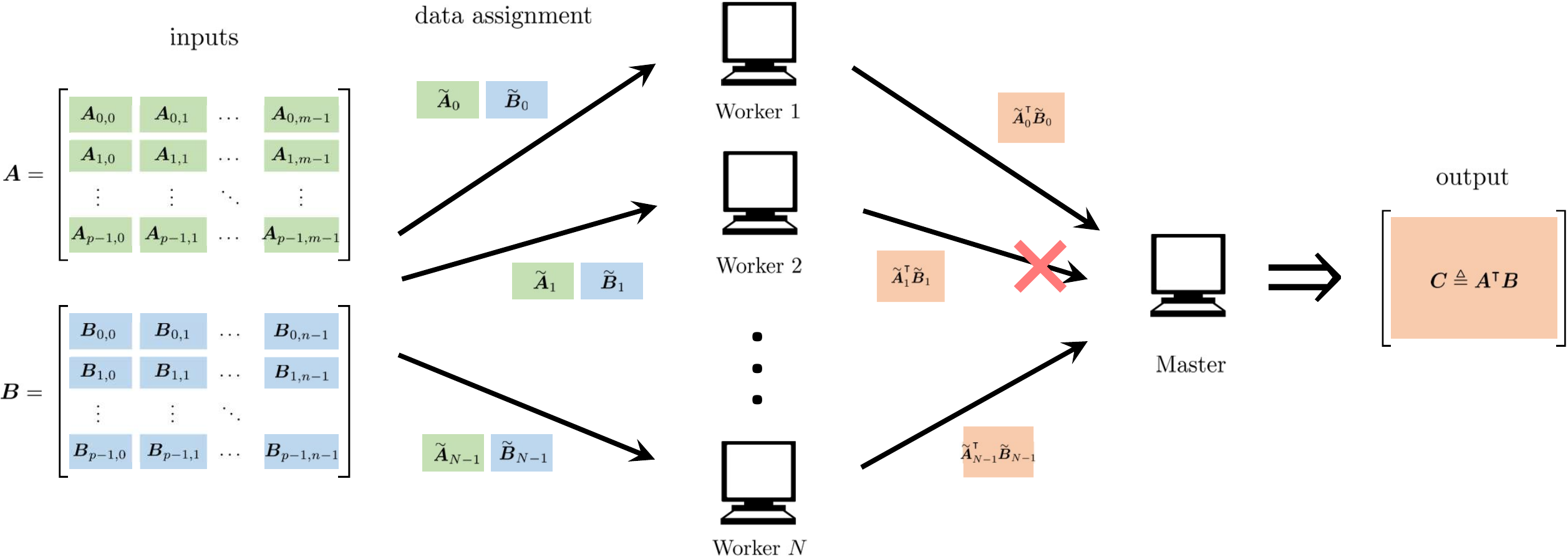}
			\caption{Overview of the distributed matrix multiplication problem with general block-partitioning of the inputs. Each worker is assigned two possibly coded submatrices and computes their product. The master aims to decode the product of the input matrices based from results from non-straggling workers.
}		%	\vspace{-3mm}
			\label{fig:sys}
		\end{figure}

An important application of PCC is to consider a more generalized matrix multiplication setting where the inputs are block-wise partitioned.
In a basic setup, given a pair of input matrices $A\in\bF^{s\times t}$, ${B}\in \bF^{s\times r}$ for a sufficiently large field $\bF$%(or two lists of matrices, to be specified later)
, each worker $i$ is assigned a pair of possibly coded matrices $\tilde{A}_i\in  \bF^{\frac{s}{p}\times \frac{t}{m}}$ and  $\tilde{B}_i\in \bF^{\frac{s}{p}\times \frac{r}{n}}$, which are encoded based on some (possibly random) functions of the input matrices respectively (see Fig. \ref{fig:sys}). The workers can each compute $\tilde{C}_i\triangleq \tilde{A}_i^\intercal \tilde{B}_i$ and return them to the master. The master tries to recover the final product $C\triangleq A^\intercal B$ based on results from a subset of fastest workers using some decoding functions. By partitioning the input matrices into $p$-by-$m$ and $p$-by-$n$ subblocks, the product of two input matrices can thus be viewed as linear combinations of $pmn$ submatrix products according to block-matrix-multiplication rules, which can be computed using $pmn$ workers with uncoded inputs.
%We say a coded computing scheme achieves a \emph{recovery threshold} of $R$, if the master can correctly decode the final output given the computing results from \emph{any} subset of $R$ workers. %\footnote
%{Note that recovery threshold is an equivalent measure of straggler resiliency, because it is identical to the difference between the number of workers and the number of stragglers that can be tolerated.}
We aim to achieve the minimum possible recovery threshold given parameters $p,m,n,$ and $N$.

%For example for block-partitioned matrix multiplication, the goal is to distributedly multiply matrices with sizes of $\bF^{t\times s}$ and $\bF^{s\times r}$, for a sufficiently large field $\bF$, %with a set of $N$ workers each 
%and each worker can multiply a pair of possibly coded matrices with sizes  $\bF^{\frac{t}{m}\times \frac{s}{p}}$ and $\bF^{\frac{s}{p}\times \frac{r}{n}}$. 

%algebraic designs 

% However for      

%3. problem fix f,g,N 
%the goal is given f  g and $N$ tolerate as many as possible. 

%1. compute over coded results. how to make sense .  
%Why encoding. 
%Encoding can help (linear computation) f(c)

%2. pcc. what is pcc. optimal codes. benifites, linear code propergates? fast interp. 
%polynomial coded computing. 

%\cite{ ucb, dot, poly, }. 

%TODO: demonstrate 

%1 para on bilinear complexity?
%For block-partitioned coded matrix multiplication, the goal is to distributedly multiply matrices with sizes of $\bF^{t\times s}$ and $\bF^{s\times r}$, for a sufficiently large field $\bF$, with a set of $N$ workers each can multiply a pair of possibly coded matrices with sizes  $\bF^{\frac{t}{m}\times \frac{s}{p}}$ and $\bF^{\frac{s}{p}\times \frac{r}{n}}$. 

%Each worker can take a possibly coded version of the input variables as input, and compute a single evaluation of some function $f$.  which can be viewed as building blocks for computing $g$. 

%\subsection{Polynomial Coded Computing}

%\subsection{Block-wise Distributed Matrix Multiplication and Entangled Polynomial Codes}

The best known recovery threshold for computing block-partitioned matrix multiplication is achieved by a class of PCC designs referred to as entangled polynomial codes. In particular, 
%PCC has been applied to compute block-partitioned matrix multiplication for straggler mitigation, and the best known result is achieved by coding designs referred to as entangled polynomial codes\cite{8437563}. 
%Explicitly, 
%The goal is to achieve the minimum possible recovery threshold given parameters $p,m,n,$ and $N$. 
%The %state of the art for straggler mitigation is
%best known entangled polynomial codes, which achieves a recovery threshold of $\min\{pmn+p-1,2R(p,m,n)-1\}$.
entangled polynomial codes achieves a recovery threshold of $\min\{pmn+p-1,2R(p,m,n)-1\}$ for any $p,m,$ and $n$\cite{8437563}. Here $R(p,m,n)$ denotes the bilinear complexity \cite{gs005} for multiplying two matrices of sizes $m$-by-$p$ and $p$-by-$n$. It is well know that $R(p,m,n)$ is subcubic, i.e, $R(p,m,n)=o(pmn)$ when $p$, $m$, and $n$ are large. Hence, it order-wise outperforms other block-partitioning based schemes in related works for straggler mitigation (e.g., \cite{8262882}).

We present entangled polynomial codes as follows. The input matrices are partitioned into $p$-by-$m$ and $p$-by-$n$ submatrices of equal sizes, denoted $A_{j,k}$ and $B_{j,k'}$ for $j\in\{0,...,p-1\}$, $k\in\{0,...,m-1\}$, and $k'\in\{0,...,n-1\}$, and we aim to recover $C_{k,k'}\triangleq \sum_{i} A_{j,k}^{\intercal}B_{j,k'}$ for any $k$ and $k'$.   
%The entangled polynomial codes partitioned the input variables into $p$-by-$m$ and $p$-by-$n$ submatrices, 
A basic version of entangled polynomial codes is designed to achieve %optimal .
%It achieves
a recovery threshold of $pmn+p-1$, which generalizes the Polynomial codes, and encodes the input variables using $c(x)=(\sum_{j=0}^{p-1}\sum_{k=0}^{m-1}  A _{j,k} x^{j+kp},\sum_{j=0}^{p-1}\sum_{k'=0}^{n-1}  B _{j,k'} x^{p-1-j+k'pm})$. Explicitly, given any distinct evaluation points $y_1,...,y_N$, each worker $i$ obtains  $(\tilde{A}_i,\tilde{B}_i)$ where  
	\begin{align}
		\tilde{A}_i&=\sum_{j=0}^{p-1}\sum_{k=0}^{m-1}  A _{j,k} y_i^{j+kp},\nonumber\\
	    \tilde{B}_i&=\sum_{j=0}^{p-1}\sum_{k'=0}^{n-1}  B _{j,k'} y_i^{p-1-j+k'pm}.
	\end{align} 
This coding construction results in the following composed polynomial 
	\begin{align}
	    f(c(x))&=\sum_{j=0}^{p-1}\sum_{k=0}^{m-1} \sum_{j'=0}^{p-1}\sum_{k'=0}^{n-1} A _{j,k}^\intercal  B _{j',k'} x^{(p-1+j-j')+kp+k'pm},
	\end{align}
	which has a degree of $pmn+p-2$, and the $mn$ needed linear combinations are exactly provided by $mn$ of its coefficients. 
	
	Importantly, in the same paper, an improved version of the entangled polynomial code is presented to approach the optimal recovery threshold for general $p,m$, and $n$, which achieves a recovery threshold of $2R(p,m,n)-1$ (Theorem 3, \cite{8437563}). Given any upper bound construction of $R(p,m,n)$ (e.g., Strassen's construction) with rank $R$ and tensor tuples $a\in\mathbb{F}^{R\times p\times m}$, $b\in \mathbb{F}^{R\times p\times n}$, and $c\in \mathbb{F}^{R\times m\times n}$, the inputs are each pre-encoded into a list of $R$ coded submatrices.\footnote{For detailed definitions, see \cite{8949560}.}  
	\begin{align}
    	\tilde{A}_{i,\textup{vec}}\triangleq\sum_{j,k} A_{j,k} a_{ijk},  \ \ \ \ 
	    \tilde{B}_{i,\textup{vec}}\triangleq\sum_{j,k} B_{j,k} b_{ijk}.
	\end{align}
Then the variables are encoded using $c(x)=\sum_j (\tilde{A}_{j,\textup{vec}}, \tilde{B}_{j,\textup{vec}} )\cdot \prod_{k\neq {j}}\frac{(x-x_k)}{(x_j-x_k)}$. I.e., each worker obtains 
	\begin{align}
    	\tilde{A}_{i}&=\sum_j \tilde{A}_{j,\textup{vec}} \cdot \prod_{k\neq {j}}\frac{(y_i-x_k)}{(x_j-x_k)}, \\ 
	    \tilde{B}_{i}&=\sum_j \tilde{B}_{j,\textup{vec}} \cdot  \prod_{k\neq {j}}\frac{(y_i-x_k)}{(x_j-x_k)},
	\end{align}
	where $x_1$,...,$x_R$ are arbitrary distinct elements of $\bF$. The coding construction provides a composed polynomial $f(c)$ with degree $2R-2$, which could achieve   $2R(p,m,n)-2$ by using upper bound constructions with a rank of $R(p,m,n)$. The final results can be decoded by re-evaluating the composed polynomial at points $x_1,...,x_R$, then combining them based on tensor $c$. 

Note that even for cases where $R(p,m,n)$ is not yet known, one can still obtain explicit coding constructions by swapping in any upper bound constructions (e.g, \cite{4567976,doi:10.1137/0120004,laderman1976noncommutative,Bini1980,Sch_nhage_1981,doi:10.1137/0211020,Coppersmith:1981:ACM:1398510.1382702,Strassen:1986:AST:1382439.1382931,COPPERSMITH1990251, stothers2010complexity,Drevet2011,Williams12multiplyingmatrices,Smirnov2013,sedoglavic2017non,sedoglavic:hal-01572046}). Subcubic recovery thresholds can still be achieved for any sufficiently large $p$, $m$, and $n$ even one only applies the well known Strassen's construction \cite{4567976}. Hence, for simplicity, in this work we present all results in terms of $R(p,m,n)$, and explicit subcubic constructions can be obtained in the same way.

We focus on linear codes, defined similarly as in  \cite{8949560, pmlr-v89-yu19b}, which guarantees linear coding complexities w.r.t. the sizes of input matrices, and are dimension independent. Precisely, in a linear coding design, the input matrix $A$ (or each input $A$ for more general settings) is partitioned into $p$-by-$m$ subblocks of equals sizes (and possibly padded with a list of i.i.d. uniformly random matrices of same sizes, referred to as random keys%, when security of the input data is taken into account
).\footnote{To make sure the setting is well defined, we assume $\bF$ is finite whenever data-security or privacy is taken into account. %, where random keys are needed.
} Matrix (or matrices) $B$ are partitioned similarly. Each worker is then assigned a pair of linear combinations of these two lists of submatrices as coded inputs. Moreover, the master uses decoding functions that computes linear combinations of received computing results.\footnote{Note that by relaxing certain assumptions made in the paper, such as 
allowing the decoder to access inputs and random keys and allowing extra computational cost at workers or master, one can further reduce the recovery threshold (e.g., see discussions in \cite{jia2019capacity, 8849245}).}

All results presented in this paper for distributed matrix multiplication directly extends to general codes with possibly non-linear constructions, by swapping any upper bound of $R(p,m,n)$ into the number workers required by any computing scheme, as we illustrated in \cite{8949560}.

\section{Secure, Private, Batch Distributed Matrix Multiplication and Main Results}\label{sec:main}

Distributed matrix multiplication is well studied in the context of straggler mitigation. Our goal in this work is to leverage entangled polynomial codes to %3 new problems. these, and show how it provides opt design
%We show that entangled polynomial codes can be adapted to
the settings of secure, private and batch distributed matrix multiplication, achieving order-wise improvement with subcubic recovery thresholds while meeting the systems' requirements. %We are going to introduce the problem and show ent as an effective way of doing them
 %The goal is to demonstrate the gain, and we focus on the second version
%\begin{remark}
%Why linear codes: 1 coding overhead. 2 Dimension independent
%\end{remark}
%We start by formally introducing the settings.

\subsection{Secure Distributed Matrix Multiplication}

%todo: break prior works into three 

Secure distributed matrix multiplication follows a similar setup discussed in Section \ref{sec:pre}, where the goal is to multiply a single pair of matrices, with additional constraints that either one or both of the input matrices are information-theoretic private to the workers, even if up to a certain number of them can collude. 
In particular, we say an encoding scheme is one-sided $T$-secure, if 
\begin{align}
    I(\{\tilde{A}_i\}_{i\in\mathcal{T}};A)=0
\end{align}
for any subset $\mathcal{T}$ with size of at most $T$, where $A$ is generated uniformly at random.  Similarly, we say an encoding scheme is fully $T$-secure, if instead 
\begin{align}
    I(\{\tilde{A}_i, \tilde{B}_i\}_{i\in\mathcal{T}};A,B )=0
\end{align}
is satisfied for any $|\mathcal{T}|\leq T$, for uniformly randomly generated $A$ and $B$.

Secure distributed matrix multiplication has been studied in \cite{chang2018capacity, 8382305, DBLP:journals/corr/abs-1810-13006, DBLP:journals/corr/abs-1812-09962, 8613446, DBLP:journals/corr/abs-1901-07705, kim2019private, 8675905 ,chang2019upload, 8761275, nodehi2019secure, jia2019capacity, kakar2019uplinkdownlink, aliasgari2019private, d2019degree}. In particular, \cite{DBLP:journals/corr/abs-1901-07705, aliasgari2019private, 8613446} presented coded computing designs for general block-wise partitioning of the input matrices, all requiring at least $pmn$ workers' computation.\footnote{In addition, at least $T$ extra workers are needed per each input matrix required to be stored securely. } 
Entangled polynomial codes achieves subcubic recovery thresholds for both one-sided and fully secure settings, formally stated in the following theorem.

\begin{theorem}\label{thm:sec}
For secure distributed matrix multiplication, there are one-sided $T$-secure linear coding schemes that achieves a recovery threshold of $2R(p,m,n)+T-1$, and fully $T$-secure linear coding schemes that achieves a recovery threshold of $2R(p,m,n)+2T-1$.  
\end{theorem}

\begin{remark}\label{rem:sec_num}
Entangled polynomial codes order-wise improves the state of the arts for general block-wise partitioning \cite{DBLP:journals/corr/abs-1901-07705, aliasgari2019private, 8613446},  by providing explicit constructions that require subcubic number of workers. In particular, the fully $T$-secure codes %can be obtained from the computing design proposed in \cite{}, they 
%are
presented in   
\cite{DBLP:journals/corr/abs-1901-07705, aliasgari2019private, 8613446} %. However,
all %these designs
require using at least $pmn+2T$ workers, while the fully $T$-secure entangled polynomial codes only requires at most $O(\frac{pmn}{\min\{p,m,n\}^{0.19}})+2T$ workers according to Strassen's upper bound, which is order-wise smaller for any large $p,m$, and $n$. 
%To demonstrate the order-wise coding gain, it suffice to upper bound the recovery threshold of the fully $T$-secure entangled polynomial codes using Strassen's construction: Note that $R(p,m,n)$ 
Similarly,  entangled polynomial codes requires order-wise smaller number of workers compared to the one-sided $1$-secure coding scheme proposed in \cite{ aliasgari2019private} (which also satisfies a privacy requirement, to be discussed in the next subsection). 
Moreover, entangled polynomial codes simultaneously handles data security and straggler issues by tolerating arbitrarily many stragglers while maintaining the same recovery threshold and privacy level. 
%all requiring at least $pmn$ workers' computation in our framework.
\end{remark}

\begin{remark}
Entangled polynomial codes enables breaking the ``cubic'' barrier when resiliency, security, or privacy is required, by providing a class of codes that operates upon any general coding structures. It supports codes developed based on ideas from matrix multiplication algorithms (which does not naturally provide resiliency or security\footnote{For example, Strassen's construction \cite{Strassen1969} leads to a computing design with $7$ workers for the most basic setting \cite{8437563}, however, one can not always achieve the same recovery threshold for the same matrix partitioning even for tolerating $1$ straggler. }), and injects tailored coding designs to allow achieving similar sub-cubic recovery properties. 

\end{remark}

\begin{remark}\label{rem:conv1}
Following similar converse proof steps we developed in \cite{8949560, pmlr-v89-yu19b}, one can show that any linear code that is either one-sided $T$-secure or fully $T$-secure requires using at least $R(p,m,n)+T$ workers. Hence, entangled polynomial codes enables achieving optimal recovery thresholds within a factor of $2$ for both settings.  
\end{remark}

\subsection{Private Distributed Matrix Multiplication}

Private distributed matrix multiplication has been studied in \cite{kim2019private,chang2019upload, aliasgari2019private, 8832193}, where the goal is to instead multiply a matrix $A$ by one of the matrices $B^{(D)}$ from $\boldsymbol{B}=(B^{(1)},..., B^{(M)})$ while keeping the request $D$ private to the workers. In particular, the master sends a (possibly random) query $Q_i$ to each worker $i$ based the request $D$. Then the matrices $\boldsymbol{B}$ are encoded by each worker $i$ into a coded submatrix $\tilde{B}_i\in \bF^{\frac{s}{p}\times \frac{r}{n}}$ based on $Q_i$. The matrix $A$ is encoded the same as the basic setting, and each worker computes the product of their coded matrices.
%the encoding function of matrices $\boldsymbol{B}$ for each worker $i$ is determined separately based on $Q_i$.
%, denoted by $g^{(Q_i)}_i$ 
%can be based on a query that the master send
%, is assigned by the master based on the index $D$.

The index $D$ should be kept private to any single worker in the sense that\footnote{Note that a stronger privacy condition $I(D;Q_i,\tilde{A}_i,A, \boldsymbol{B})=0$ can still be achieved, if one use the scheme for private and secure distributed matrix multiplication presented later in this paper. } 
%In , the authors considered a scenario where the computation task is 
\begin{align}
I(D;Q_i,\tilde{A}_i,\boldsymbol{B})=0    
\end{align}
for any $i\in[N]$, where $A,\boldsymbol{B},D$ are sampled uniformly at random. The master can decode the final output based on the returned results, the request $D$ and query $Q_i$'s.

Moreover, in some related works \cite{aliasgari2019private, kim2019private, chang2019upload}, the encoding of $A$ is also required to be secure against any single curious worker. I.e., 
\begin{align}
I(\tilde{A}_i;A)=0    
\end{align}
for any $i\in[N]$ if $A$ is sampled uniformly random. This setting is referred to as private and secure distributed matrix multiplication. 

The state of the art for private and secure distributed matrix multiplication with general block-partitioning based designs was proposed in \cite{aliasgari2019private}, which requires at least $pmn$ number of workers. Entangled polynomial codes achieves subcubic recovery thresholds, formally stated in the following theorem.

%studied in blah, what is achieved

\begin{theorem}\label{thm:pri}
For private coded matrix multiplication, there are linear coding schemes that achieve a recovery threshold of $2R(p,m,n)$. For private and secure distributed matrix multiplication, linear coding schemes can achieve a recovery threshold of $2R(p,m,n)+1$.  
\end{theorem}

\begin{remark}
Entangled polynomial codes order-wise improves the state of the arts for general block-wise partitioning \cite{aliasgari2019private}, by providing explicit constructions that achieves subcubic recovery thresholds, while simultaneously provides straggler-resiliency, data-security and privacy. As mentioned in Remark
\ref{rem:sec_num}, \cite{aliasgari2019private} presents a private and secure matrix multiplication design that requires at least $pmn+1$ workers, while entangled polynomial codes provides a private and secure design that requires at most $O(\frac{pmn}{\min\{p,m,n\}^{0.19}})+1$ workers according to Strassen's upper bound, which is order-wise smaller for any large $p,m$, and $n$.
%all requiring at least $pmn$ workers' computation in our framework.  

\end{remark}

\begin{remark}
Similar to the discussion in Remark \ref{rem:conv1}, one can show that any linear code requires at least $R(p,m,n)$ workers for private coded matrix multiplication and $R(p,m,n)+1$ workers for private and secure distributed matrix multiplication, even if one ignore the privacy requirement. This indicates a factor-of-$2$ optimality  of entangled polynomial codes for both settings.
\end{remark}

Entangled polynomial codes also applies to a more general scenario where the encoding functions for both input matrices are assigned to the workers, which we refer to as \emph{fully private coded matrix multiplication} and formulate as follows. In fully private coded matrix multiplication, we have two lists of input matrices $\boldsymbol{A}=(A^{(1)},..., A^{(M)})$ and $\boldsymbol{B}=(B^{(1)},..., B^{(M)})$, and the master aims to compute ${A^{(D)}}^\intercal B^{(D)}$ given an index $D$. We assume $M>1$, because otherwise the privacy requirement is trivial.

We aim to find computation designs such that $D$ is private against any single workers. Explicitly, the master sends a (possibly random) query $Q_i$ to each worker $i$ based on the demand $D$. Then worker $i$ encodes %encoding functions for
\emph{both} %matrices
$A$ and $B$ based on $Q_i$. We require the requests to be private in the sense that 
\begin{align}
I(D; Q_i,\boldsymbol{A},\boldsymbol{B})=0    
\end{align}
for any $i\in[N]$, where $\boldsymbol{A},\boldsymbol{B},D$ are sampled uniformly at random.
%\footnote{%Note that any coded computing scheme satisfying this condition 
%Our coded computing scheme also applies if $(D_A, D_B)$ is instead sampled from any subset of possible values.}
We summarize the performance of entangled polynomial codes for fully private coded matrix multiplication as follows.

\begin{theorem}\label{thm:fpri}
For fully private coded matrix multiplication, there are linear coding schemes that achieve a recovery threshold of $2R(p,m,n)+1$.  
\end{theorem}

\begin{remark}
Similar to earlier discussions, entangled polynomial codes provides coding constructions for fully private coded matrix multiplication with subcubic recovery thresholds. 
One can prove that any fully private linear code requires at least $R(p,m,n)+1$ workers. 
%Because any linear code requires at least $R(p,m,n)$ workers. 
Hence, the factor-of-$2$ optimality of entangled polynomial codes also holds true for fully private coded matrix multiplication.
\end{remark}

\subsection{Batch Distributed Matrix Multiplication}\label{subsec:m_bathc}

The authors of \cite{jia2019capacity, jia2019cross, jia2019generalized} considered a scenario where the goal is to compute $L$ copies of the matrix multiplication task in one round of communication. Formally, a basic setting for batch distributed matrix multiplication is that we have two lists of input matrices $\boldsymbol{A}=(A^{(1)},..., A^{(L)})$ and $\boldsymbol{B}=(B^{(1)},..., B^{(L)})$, and the master aims to compute their element-wise product $\boldsymbol{C}=({A^{(1)}}^\intercal B^{(1)},..., {A^{(L)}}^\intercal B^{(L)})$. Given partitioning parameters $p,m,$ and $n$, each worker still computes a single multiplication of coded submatrices with sizes  $\bF^{\frac{t}{m}\times \frac{s}{p}}$ and $\bF^{\frac{s}{p}\times \frac{r}{n}}$.  

%These works focus on reducing recovery thresholds and no security or privacy is required. 

For general block-partitioning based schemes, the state of the art design is provided in \cite{jia2019cross, jia2019generalized}, where the focus is to reduce the recovery threshold and no security or privacy is required. All known coding constructions presented for batch distributed matrix multiplication requires cubic number of workers per each multiplication task even no straggler presents (i.e., requiring at least $Lpmn$ workers in total).  

We show that %entangled polynomial codes provides order-wise improvement for batch matrix multiplication. Moreover, 
entangled polynomial codes offer a unified coding framework for batch matrix multiplication, achieving subcubic recovery thresholds while simultaneously handling all security and privacy requirements that are discussed earlier in this section. We present this result in the following theorem.\footnote{Similar to \cite{8949560}, in the most basic scenario with no requirements on resiliency, security, and privacy (i.e., requiring a recovery threshold of $N$, with $T=0$ and $M=1$), one can directly apply any upper bound construction of bilinear complexity for batch matrix multiplication to further reduce the number of workers by a factor of $2$. However, here we focus on demonstrating the coding gain and present the results for general scenarios.}  The proofs and detailed formulations can be found in Section \ref{sec:batch}.

%studied in blah, what is achieved

%Distributed matrix multiplication is recently  studied in the context of providing 
%security against eavesdropping workers \cite{chang2018capacity, 8382305, DBLP:journals/corr/abs-1810-13006, DBLP:journals/corr/abs-1812-09962, 8613446, DBLP:journals/corr/abs-1901-07705, kim2019private, 8675905 ,chang2019upload, 8761275, nodehi2019secure, jia2019capacity, kakar2019uplinkdownlink, aliasgari2019private, d2019degree}, privacy of request \cite{kim2019private,chang2019upload, aliasgari2019private, 8832193}, and supporting batch evaluation \cite{jia2019capacity, jia2019cross, jia2019generalized}. Block-partitioning based designs has been proposed in \cite{DBLP:journals/corr/abs-1901-07705, aliasgari2019private, 8613446, jia2019cross, jia2019generalized}, for these problems, all requiring at least cubic (i.e., $pmn$) number of workers' computation. 

\begin{theorem}\label{thm:batch}
For coded distributed batch matrix multiplication with parameters $p,m,n,$ and $L$, there are linear coding schemes that achieve a recovery threshold of $2LR(p,m,n)-1$. Moreover, for extended settings in batch matrix multiplication, linear coding schemes achieve the following recovery thresholds:
\begin{itemize}
    \item One sided $T$-security: $2LR(p,m,n)+T-1$,
    \item Fully $T$-security: $2LR(p,m,n)+2T-1$,
    \item Privacy of request: $2LR(p,m,n)$,
    \item Security and Privacy: $2LR(p,m,n)+1$,
    \item Full Privacy: $2LR(p,m,n)+1$.
\end{itemize}
\end{theorem}

\begin{remark}
Entangled polynomial codes provide coding schemes that order-wise improves the state-of-the-art schemes in \cite{jia2019cross, jia2019generalized} for batch matrix multiplication when the matrices are block-wise partitioned. The coding designs proposed in \cite{jia2019cross, jia2019generalized} focused on straggler mitigation and requires a recovery threshold of at least $Lpmn$, with computation and storage costs equal or greater than the framework considered in this paper; while entangled polynomial codes achieves order-wise smaller recovery thresholds for any large $p$, $m$ and $n$.  
\end{remark}

\begin{remark}
Note that batch-multiplying $L$ pairs of matrices is still computing a bilinear function, one can simply use similar bilinear decomposition bounds for this operation as in \cite{8949560}, and all earlier achievability and converse results extended to batch computation. 
However, to better demonstrate the achievability of subcubic recovery thresolds, we present our results based on a subadditivity upper bound.\footnote{Specifically, let $R(L,p,m,n)$ denote the bilinear complexity of batch multiplying $L$ pairs of $m$-by-$p$ and $p$-by-$n$ matrices. We have $R(L,p,m,n)\leq L R(p,m,n)$.}  One can similarly prove the factor-of-$2$ optimalities for the general entangled polynomial codes framework for all settings we presented for batch matrix multiplication.  
%We state the corresponding results in the following theorem, for detailed formulations, see Section \ref{sec:batch}. 
\end{remark}

%[] considers a general framework where each worker stores and computes more? Because LCC naturally extended to any bilinear, we also blah, see remark blah wrong???

%we can still find the tensor rank, but as an upper bound. 

%Note, if they can compute in parallel, this is the same. for brevity

%Our work connect to batch, which natrually extends to scenarios in . same comp comm

%with flexible partitioning

%Polynomial Coded Computing (PCC) framework. Extends to general computation. Includes works such as. 

%pairwise, convolution, inner, elementwise prod, and batch poly eval. Block is first in blah

%Approach to Privacy Preserved Data Mining in Distributed Systems

%papers in ISIT

%and papers that cites them

%check all papers that cites polydot or generalized polydot

%eavesdropping and colluding workers 

%secure \cite{chang2018capacity, 8382305, DBLP:journals/corr/abs-1810-13006, DBLP:journals/corr/abs-1812-09962, 8613446, DBLP:journals/corr/abs-1901-07705, kim2019private, 8675905 ,chang2019upload, 8761275, nodehi2019secure, jia2019capacity, kakar2019uplinkdownlink, aliasgari2019private, d2019degree}, private \cite{kim2019private,chang2019upload, aliasgari2019private, 8832193}, batch \cite{jia2019capacity, jia2019cross, jia2019generalized}

%block \cite{DBLP:journals/corr/abs-1901-07705, aliasgari2019private, 8613446, jia2019cross, jia2019generalized}

\section{Achievability Schemes for Secure Distributed Matrix Multiplication} \label{sec:sec}

%To achieve any $R\geq R(p,m,n)$ 

In this section, we present coding schemes for the simple scenario where the only additional requirement for distributed matrix multiplication is to maintain the security of input matrices. This provides a proof for Theorem \ref{thm:sec}.

Given parameters $p$, $m$, and $n$, we denote the partitioned uncoded input matrices by $\{A_{i,j}\}_{i\in [p], j\in[m]}$ and $\{B_{i,j}\}_{i\in [p], j\in[n]}$. The encoding  consists of two steps.

In Step 1, given any upper bound construction of $R(p,m,n)$ (e.g., Strassen's construction) with rank $R$ and tensor tuples $a\in\mathbb{F}^{R\times p\times m}$, $b\in \mathbb{F}^{R\times p\times n}$, and $c\in \mathbb{F}^{R\times m\times n}$, we pre-encode the inputs each into a list of $R$ coded submatrices.\footnote{For detailed definitions of bilinear complexity and upper bound constructions, see \cite{8949560}.}  
	\begin{align}
    	\tilde{A}_{i,\textup{vec}}\triangleq\sum_{j,k} A_{j,k} a_{ijk},  \ \ \ \ 
	    \tilde{B}_{i,\textup{vec}}\triangleq\sum_{j,k} B_{j,k} b_{ijk}.
	\end{align}
As we have explained in \cite{8949560}, this pre-encoding essentially provides a linear coding scheme with $R$ workers that does not provide straggler-resiliency and data-security, of which we need to take into account in the second part of the encoding.        

In Step 2, note that it suffice to recover the element-wise products $\tilde{A}_{1,\textup{vec}}^\intercal \tilde{B}_{1,\textup{vec}},...,\tilde{A}_{R,\textup{vec}}^\intercal \tilde{B}_{R,\textup{vec}}$. We can build upon optimal coding constructions for element-wise multiplication, first presented in \cite{8437563} for straggler mitigation and then extended in \cite{pmlr-v89-yu19b} to also provide data-privacy. 

We first pad the two vectors $\{\tilde{A}_{i,\textup{vec}}\}_{i\in[R]}$ and   $\{\tilde{B}_{i,\textup{vec}}\}_{i\in[R]}$ with uniformly random keys. If matrix $A$ needs to be stored securely against up to $T$ colluding workers, we pad the pre-coded matrices of $A$ with $T$ uniformly random matrices $Z_{1},...,Z_{T}\in\bF^{\frac{s}{p}\times\frac{t}{m}}$. Explicitly, we define
\begin{align}
    \boldsymbol{\tilde{A}}_{\textup{vec}}'\triangleq ({\tilde{A}_{1,\textup{vec}}},...,{\tilde{A}_{R,\textup{vec}}},Z_{1},...,Z_{T})
\end{align}
      if $A$ needs to be stored securely; otherwise, we define 
    \begin{align}
    \boldsymbol{\tilde{A}}_{\textup{vec}}'\triangleq ({\tilde{A}_{1,\textup{vec}}},...,{\tilde{A}_{R,\textup{vec}}}).
\end{align}
Similarly, we define vector  $\boldsymbol{\tilde{B}}_{\textup{vec}}'$ for matrix $B$ in the same way. For brevity, we denote the lengths of $\boldsymbol{\tilde{A}}_{\textup{vec}}'$ and $\boldsymbol{\tilde{B}}_{\textup{vec}}'$ by $L_A$ and $L_B$.

Then we arbitrarily select $R+T$ distinct elements from $\bF$, denoted $x_1,...,x_{R+T}$, and $N$ distinct elements from $\bF\backslash \{x_1,...,x_{R}\}$, denoted $y_1,...,y_{N}$. We encode the inputs for each worker $i$ as follows.
	\begin{align}
    	\tilde{A}_{i}&=\sum_{j\in[L_A]} \tilde{A}_{j,\textup{vec}}' \cdot \prod_{k\in[L_A]\backslash {j}}\frac{(y_i-x_k)}{(x_j-x_k)}, \\ 
	    \tilde{B}_{i}&=\sum_{j\in[L_B]} \tilde{B}_{j,\textup{vec}}' \cdot  \prod_{k\in[L_B]\backslash {j}}\frac{(y_i-x_k)}{(x_j-x_k)}.
	\end{align}
As proved in \cite{pmlr-v89-yu19b}, the above encoding scheme satisfies the requirements for both one-sided and fully $T$-secure settings.\footnote{Such property is referred to as $T$-private in \cite{10.1145/62212.62213, pmlr-v89-yu19b}}

According to the PCC framework, we have encoded the input matrices using polynomials with degrees of $L_A-1$ and  $L_B-1$, where each worker $i$ is assigned their evaluations at $y_i$. Hence, after the workers multiply their coded matrices, they obtain evaluations of the multiplicative product of these polynomials, which has degree $L_A+L_B-2$. Note that evaluations of this composed polynomial at  $x_1,...,x_{R}$ recovers the needed element-wise products. The decodability requirement of PCC is satisfied.    
Consequently, the master can recover the final output by interpolating the composed polynomial after sufficiently many results from the workers are received, achieving a recovery threshold of $L_A+L_B-1$.  

Recall that for one-sided $T$-secure setting, we have $L_A=R+T$ and $L_B=R$; then for fully $T$-secure setting, we have $L_A=L_B=R+T$. Hence, we have obtained linear coding schemes with recovery thresholds of $2R+T-1$ and $2R+2T-1$ for both settings respectively given any upper bound constructions of $R(p,m,n)$ with rank $R$. Fundamentally, there exists constructions that exactly achieves the rank $R(p,m,n)$, which proves the existance of coding schemes stated in Theorem \ref{thm:sec}.

\begin{remark}
The coding scheme we presented for computing element-wise product with one-sided privacy naturally extends to provide optimal  codes for the scenario of batch computation of multilinear functions where each of the input entries are coded to satisfy possibly different security requirements.

%We present the following corollary, of which the detailed setup and proof are presented in Appendix \ref{app:ss_mul}. %\footnote{The basic case $(T_1$,...,$T_d)=\boldsymbol{0}$ is already covered in \cite{pmlr-v89-yu19b}. Hence, we focus on cases where the security requirement is non-trivial.} 
\end{remark}

%\begin{corollary}
%Consider a scenario of batch computing any multilinear function $F$ of degree $d$ given a list of $R$ inputs, where each worker can compute a single evaluation of $f$ over separately coded entries with $T_1$,...,$T_d$-security. When at least one of $T_i's$ is non-zero, the optimal linear code achieves a recovery threshold of $d(R-1)+1+\sum_{i}T_i$, which also equals the minimum possible number of workers required for any linear code. 
%\end{corollary}

%The coding scheme we had for computing element-wise product with full privacy naturally extends to the scenario of batch computation of multilinear functions where each of the input entries are coded to satisfy with possibly different security requirements. We present the following corollary, of which the detailed setup and proof are presented in Appendix \ref{app:ss_mul}. \footnote{The basic case $(T_1$,...,$T_d)=\boldsymbol{0}$ is already covered in \cite{pmlr-v89-yu19b}. Hence, we focus on cases where the security requirement is non-trivial.} 

%\cite{chang2018capacity}, \cite{8675905} , \cite{DBLP:journals/corr/abs-1812-09962}, %\cite{DBLP:journals/corr/abs-1810-13006}, \cite{d2019degree,  8382305} SDMM with column
%\cite{nodehi2019secure} SMPC with column
%\cite{8761275} SDMM two-sided column
%\cite{kim2019private} column private and secure

%\cite{DBLP:journals/corr/abs-1901-07705} uses block
%\cite{kakar2019uplinkdownlink} uses column
%\cite{aliasgari2019private} uses block partitioning. They also discussed a version including privacy requirements.

%\cite{8613446} also uses block partitioning. 

\section{Achievability Schemes for Private Distributed Matrix Multiplication}

In this section, we present the coding scheme for proving Theorem \ref{thm:pri} and \ref{thm:fpri}. We start with the setting for Theorem \ref{thm:pri}, where the goal is to multiply matrix $A$ by one of the matrices from $B^{(1)},...,B^{(M)}$.

Similar to Section \ref{sec:sec}, we first pre-encode the input matrices into lists of vectors of length $R$,  given any upper bound construction of $R(p,m,n)$ with rank $R$ and tensor tuples $a\in\mathbb{F}^{R\times p\times m}$, $b\in \mathbb{F}^{R\times p\times n}$, and $c\in \mathbb{F}^{R\times m\times n}$. In particular, 
given parameters $p$, $m$, and $n$, we denote the partitioned uncoded input matrices by $\{A_{i,j}\}_{i\in [p], j\in[m]}$ and $\{B_{i,j}^{(\ell)}\}_{i\in [p], j\in[n],\ell\in[M]}$. %The encoding  consists of two steps.
We define
	\begin{align}
    	\tilde{A}_{i,\textup{vec}}\triangleq\sum_{j,k} A_{j,k} a_{ijk},  \ \ \ \ 
	    \tilde{B}^{(\ell)}_{i,\textup{vec}}\triangleq\sum_{j,k} B_{j,k}^{(\ell)} b_{ijk},
	\end{align}
for each $i\in [R]$ and $\ell\in[M]$. Then given any request $D\in[M]$, it suffices to compute the element-wise product $\{\tilde{A}_{i,\textup{vec}}^\intercal \tilde{B}^{(D)}_{i,\textup{vec}}\}_{i\in[R]}$ while keeping $D$ private.    

The second part of the encoding scheme is motivated by coding ideas developed in \cite{8437563, kim2019private} and earlier sections. In particular, we first pad the pre-encoded vector of $A$ with random keys for security. We define 
\begin{align}
    \boldsymbol{\tilde{A}}_{\textup{vec}}'\triangleq ({\tilde{A}_{1,\textup{vec}}},...,{\tilde{A}_{R,\textup{vec}}},Z),
\end{align}
if $A$ needs to be stored securely, where $Z$ is a random key sampled from $\bF^{\frac{s}{p}\times \frac{t}{m}}$ with a uniform distribution; otherwise, 
   \begin{align}
    \boldsymbol{\tilde{A}}_{\textup{vec}}'\triangleq ({\tilde{A}_{1,\textup{vec}}},...,{\tilde{A}_{R,\textup{vec}}}).
\end{align}
For brevity, we denote the length of $\boldsymbol{\tilde{A}}_{\textup{vec}}'$ by $L_A$.

We arbitrarily select $R+1$ distinct elements from $\bF$, denoted $x_1,...,x_{R+1}$, and encode matrix $A$ %similar to \cite{8437563} and Section \ref{sec:sec}
by defining the following Lagrange polynomial,
	\begin{align}
    	\tilde{A}(x)&\triangleq \sum_{i\in[L_A]} \tilde{A}'_{i,\textup{vec}} \cdot \prod_{j\in[L_A]\backslash {i}}\frac{(x-x_j)}{(x_i-x_j)},
	\end{align}
	  %; otherwise, we use the same Lagrange expression with 
We then arbitrarily select a finite subset $\mathcal{Y}$ of $\bF\backslash \{x_1,...,x_{R}\}$ with at least $N$ elements, and let the master uniformly randomly generate $N$ distinct elements from $\mathcal{Y}$, denoted $y_1,...,y_N$. The master sends  $\tilde{A}_i=\tilde{A}(y_i)$ to each worker $i$, which satisfied the security of $A$ when required.    

Given a request $D$, we similarly define
	\begin{align}
    	\tilde{B}(x)&\triangleq \sum_{i\in[R+1]} \tilde{B}'_{i,\textup{vec}} \cdot \prod_{j\in [R+1]\backslash {i}}\frac{(x-x_j)}{(x_i-x_j)},
	\end{align}
where 
\begin{align}
    \boldsymbol{\tilde{B}}'_{\textup{vec}}\triangleq ({\tilde{B}^{(D)}_{1,\textup{vec}}},...,{\tilde{B}^{(D)}_{R,\textup{vec}}},Y),
\end{align}
and $Y\in \bF^{\frac{s}{p}\times \frac{r}{n}}$ is a quantity to be specified later. If the encoding can be designed such that each worker essentially computes $\tilde{A}^\intercal(y_i) \tilde{B}(y_i)$, then we can achieve the recovery thresholds stated in Theorem \ref{thm:pri}.

To construct a private computing scheme where $\tilde{B}_i$ is equivalent to $\tilde{B}(y_i)$, we divide $\tilde{B}(x)$ by a scalar\footnote{Note that here we are exploiting the fact that each worker computes a function that is multilinear. For more general scenarios (e.g., general polynomial evaluations we considered in \cite{pmlr-v89-yu19b}), scaling the coded variables could affect decodability. } 
\begin{align}\label{eq:cx}
c(x)\triangleq \prod_{j\in [R]}\frac{(x-x_j)}{(x_{R+1}-x_j)},
\end{align}
so that the result can be expressed as the unweighted sum of $Y$ and  $\tilde{B}^{(D)}_{\textup{Norm}}(x)$ with function $\tilde{B}^{(\cdot)}_{\textup{Norm}}(x)$ defined as follows
\begin{align}
    	\tilde{B}^{(k)}_{\textup{Norm}}(x)&\triangleq -\sum_{i\in[R]} \tilde{B}^{(k)}_{i,\textup{vec}} %\cdot
    	\left(  \prod_{j\in[R]\backslash {i}}\frac{(x_{R+1}-x_j)}{(x_i-x_j)}\right)\frac{(x-x_{R+1})}{(x-x_{i})}.\nonumber
\end{align}
We let the master generate i.i.d. uniformly random variables $\{z_i\}_{i\in [M]\backslash D}$ from $\mathcal{Y}$ independent of $y_i$'s. The master sends a query $Q_i=(q_{i1},...,q_{iM})$ to each worker $i$ with $q_{ij}=y_i$ for $j=D$ and $q_{ij}=z_j$ for $j\neq D$. Because each $Q_i$ appears uniformly random to worker $i$, the presented coding scheme satisfies the privacy requirement.    

We let each worker $i$ encode $\boldsymbol{B}$ by computing $\sum_j \tilde{B}^{(j)}_{\textup{Norm}}(q_{ij})$. Consequently, each encoded variable  can be re-expressed as 
\begin{align}
    \tilde{B}_i=\frac{\tilde{B}(y_i)}{c(y_i)} 
\end{align}
with $Y=\sum_{j\neq D} \tilde{B}^{(j)}_{\textup{Norm}}(q_{ij})$ independent of $y_i$.

After the workers multiply the coded matrices, each worker $i$ essentially returns $\tilde{A}^\intercal(y_i) {\tilde{B}(y_i)}/{c(y_i)}$. Because $y_i$ is available at the  decoder, the master can decode  $\tilde{A}^\intercal(y_i) {\tilde{B}(y_i)}$ given each worker $i$'s returned result by computing $c(y_i)$. Hence, by receiving results from sufficiently workers, the master can recover the needed element-wise product by Lagrange interpolating the polynomial $\tilde{A}^\intercal(x) {\tilde{B}(x)}$, and proceed to compute the final output.

Because the degree of  $\tilde{A}^\intercal(x) {\tilde{B}(x)}$ equals $L_A-1+R$, the presented coding scheme achieves a recovery threshold of $L_A+R$. Note that $L_A=R$ when no security is required and  $L_A=R+1$ when $A$ is stored securely. We have obtained linear coding schemes with recovery thresholds of $2R-1$ for private coded matrix multiplication, and $2R$ for private and secure distributed matrix multiplication for any upperbound construction of $R(p,m,n)$, which completes the proof for Theorem \ref{thm:pri}.   

\begin{remark}
This coding scheme naturally extends to the scenario where the encoding of $A$ is required to be $T$-secure. A recovery threshold of $2R(p,m,n)+T$ can be achieved, which is optimal within a factor of $2$.
\end{remark}

We now present the coding scheme for the fully private setting. The matrices are pre-encoded the same way and we denote the corresponding matrices by   $\{  \tilde{A}^{(\ell)}_{i,\textup{vec}}, \tilde{B}^{(\ell)}_{i,\textup{vec}}\}_{i\in[R], \ell\in[M]}$. To recover the final output, it suffices to compute $\{  {\tilde{A}^{(D)\intercal}}_{i,\textup{vec}} \tilde{B}^{(D)}_{i,\textup{vec}}\}_{i\in[R]}$. 

We arbitrarily select $R+1$ distinct elements from $\bF$, denoted $x_1,...,x_{R+1}$, and define the following functions
\begin{align}
	\tilde{A}^{(k)}_{\textup{Norm}}(x)&\triangleq -\sum_{i\in[R]} \tilde{A}^{(k)}_{i,\textup{vec}}  \left( \prod_{j\in[R]\backslash {i}}\frac{(x_{R+1}-x_j)}{(x_i-x_j)}\right)\frac{(x-x_{R+1})}{(x-x_{i})}.\nonumber \\
    	\tilde{B}^{(k)}_{\textup{Norm}}(x)&\triangleq -\sum_{i\in[R]} \tilde{B}^{(k)}_{i,\textup{vec}}  \left( \prod_{j\in[R]\backslash {i}}\frac{(x_{R+1}-x_j)}{(x_i-x_j)}\right)\frac{(x-x_{R+1})}{(x-x_{i})}.\nonumber
\end{align}

We then arbitrarily select a finite subset $\mathcal{Y}$ of $\bF\backslash \{x_1,...,x_{R}\}$ with at least $N$ elements. Let the master uniformly randomly generate $N$ distinct elements from $\mathcal{Y}$, denoted $y_1,...,y_N$, and i.i.d. uniformly random variables $\{z_i\}_{i\in [M]\backslash D}$ from $\mathcal{Y}$ independent of $y_i$'s. %The master uniformly randomly shuffles $\{z_i\}_{i\in [M]\backslash D_A}$ and we denote the new sequence by $\{z'_i\}_{i\in [M]\backslash D_B}$. 
The master sends a query $Q_i=(q_{i1},...,q_{iM})$ to each worker $i$ with $q_{ij}=y_i$ for $j=D$, and $q_{ij}=z_j$ for $j\neq D$. This query is fully private, because for each worker $i$, $q_{i1},...,q_{iM}$ appears i.i.d. uniformly random in $\mathcal{Y}$. % and $q'_{i1},...,q'_{iM}$ appears to be its uniformly randomly shuffled version.

Each worker $i$ encodes the input matrices as follows
\begin{align}
   \tilde{A}_i&= \sum_j \tilde{A}^{(j)}_{\textup{Norm}}(q_{ij}),\\
   \tilde{B}_i&= \sum_j \tilde{B}^{(j)}_{\textup{Norm}}(q_{ij}).
\end{align}
After the computation result is received from any worker $i$, by multiplying a scalar factor $c^2(y_i)$ with function $c$ defined in equation (\ref{eq:cx}), the master recovers the evaluation of the product of two Lagrange polynomials of degree $R$ at point $y_i$. By interpolating this polynomial and re-evaluating it at $x_i$'s, the master can recover all needed element-wise products. This provides a coding scheme that proves Theorem \ref{thm:fpri}.

\section{Achievability Schemes for Batch Distributed Matrix Multiplication}\label{sec:batch}

In this section, we present the coding scheme for proving Theorem \ref{thm:batch}. We start with the basic setting where no security or privacy is required. As mentioned in Section \ref{subsec:m_bathc}, one can directly decompose the tensor characterizing the $L$-batch matrix multiplication, and all earlier results as well as Theorem 3 in \cite{8437563} extends to batch distributed matrix multiplication. However, we instead present one certain class of upper bounds based on subadditivity of tensor rank. 

 Explicitly, we denote the partitioned uncoded input matrices by $\{A^{(k)}_{i,j}\}_{i\in [p], j\in[m], k\in[L]}$ and $\{B^{(k)}_{i,j}\}_{i\in [p], j\in[n], k\in[L]}$.
Given any upper bound construction of $R(p,m,n)$ with rank $R$ and tensor tuples $a\in\mathbb{F}^{R\times p\times m}$, $b\in \mathbb{F}^{R\times p\times n}$, and $c\in \mathbb{F}^{R\times m\times n}$, we define
	\begin{align}
    	\tilde{A}_{i,\ell,\textup{vec}}\triangleq\sum_{j,k} A_{j,k}^{(\ell)} a_{ijk},  \ \ \ \ 
	    \tilde{B}_{i,\ell,\textup{vec}}\triangleq\sum_{j,k} B_{j,k}^{(\ell)} b_{ijk}.
	\end{align}
for each $i\in [R]$ and $\ell\in[L]$. Note that the batch product can be recovered from the element-wise product $\{\tilde{A}_{i,\ell,\textup{vec}}^\intercal \tilde{B}_{i,\ell,\textup{vec}}\}_{i\in[R],\ell\in[L]}$. One can directly apply the optimal coding scheme presented in \cite{8437563}, which encodes the pre-encoded vectors using Lagrange polynomials. According to Corollary 1 in \cite{8437563}, the resulting scheme achieves a recovery threshold of $2LR-1$, which proves the based scenario for Theorem \ref{thm:batch}.

%\cite{jia2019cross} (shorter \cite{jia2019generalized}) presented CSA code for batch matrix multiplication.
%\cite{jia2019capacity} focused on tradeoff between security and privacy, also batch?
%mohammad LCC cited

%\section{Results}
%Our result improves the recovery threshold, or the number of workers required to achieve the same security requirement. It also automatically applies to the batch evaluation setting considered in \cite{jia2019capacity, jia2019cross}.

%\section{Conclusion}

%In this paper, we demonstrated that the coding gain and optimality achieved by entangled polynomial codes can be extended to several important settings in coded distributed matrix multiplication, providing explicit constructions with order-wise improved (subcubic) recovery threshold. 

%One interesting following direction is to apply coding and converse bounding techniques to resolve more general problems in coded computing. For example, as an intermediate step for developing entangled polynomial codes in \cite{8437563}, we designed an optimal code for computing element-wise product, which encodes the input variables using Lagrange polynomial. This result is later extended in \cite{pmlr-v89-yu19b} where we prove that the same encoding is in fact optimal for computing general multivariate polynomials.
\begin{remark}
In \cite{dutta18on},  Lagrange encoding is also applied to compute inner product (sum of element-wise products) to achieve the same recovery threshold. Remarkably, \cite{dutta18on} pointed out that the encoding can be made systematic as Lagrange polynomials passes through all uncoded inputs, %by letting a subset of workers following the same steps
as stated in \cite{6820791}. 
It is mentioned in \cite{dutta18on} that the main benefits of using systematic encoding designs is to enable recovery from results of a certain smaller subset of ``systematic'' workers, %as if no ``parity workers'' exists
 which provides backward-compatibility %with designs that does not handle stragglers, 
and potentially reduces computation and decoding latency. % when the results from all ``systematic workers'' are received earlier.
Based on this observation, the entangled polynomial codes can be adapted to a ``systematic'' version that goes beyond inner product and handles generalized block-wise partitioned matrices by choosing the same evaluation points as in \cite{6820791}, so that a subset of $R(p,m,n)$ workers computes all needed ``uncoded'' products of the pre-encoded matrices, and all major benefits of systematic encoding are provided. 
 %by encoding upon any non-straggler mitigating scheme using  systematic Lagrange polynomials, the master can always start decoding the final output as long as receiving results from a smaller subset of ``systematic'' workers as if no extra coding is happening. 
This construction gives a practical solution to an open problem stated in \cite{8849811}, in the sense of achieving all major benefits of systematic encoding, and improving recovery thresholds for any sufficiently large values of $p$, $m$, and $n$.  
\end{remark}

Now we formally state the settings with security and privacy requirements. Similar to Section \ref{sec:main}, for batch matrix multiplication with security requirement, the formulation is the same as the basic setup for batch distributed matrix multiplication, except that the inputs needs to be stored information-theoretic privately even if up to $T$ workers collude. We say a coding scheme is one-sided $T$-secure,  if 
\begin{align}
    I(\{\tilde{A}_i\}_{i\in\mathcal{T}};\boldsymbol{A})=0
\end{align}
for any subset $\mathcal{T}$ with size of at most $T$, where $A$ is generated uniformly at random.  We say an encoding scheme is fully $T$-secure, if instead 
\begin{align}
    I(\{\tilde{A}_i, \tilde{B}_i\}_{i\in\mathcal{T}};\boldsymbol{A}, \boldsymbol{B} )=0
\end{align}
is satisfied for any $|\mathcal{T}|\leq T$, for uniformly randomly generated $A$ and $B$.

When privacy is taken into account, the goal is to instead batch multiply a list of $L$ matrices by one unknown subset of $L$ matrices $\boldsymbol{B}=\{B^{(i,D)}\}_{i\in[L]}$ from a set  $\boldsymbol{B}=\{B^{(i,j)}\}_{i\in[L],j\in[M]}$ while keeping the request $D$ private to the workers. The master sends a query and a coded version of $\boldsymbol{A}$ with size $\bF^{\frac{s}{p}\times \frac{t}{m}}$ to each worker,
then each worker encodes matrices $\boldsymbol{B}$ into a coded submatrix of size $\bF^{\frac{s}{p}\times \frac{r}{n}}$ based on the query, the same as in private distributed matrix multiplication. We say a computing scheme for batch matrix multiplication is private, if 
%the encoding function of matrices $\boldsymbol{B}$ for each worker $i$ is determined separately based on $Q_i$.
%, denoted by $g^{(Q_i)}_i$ 
%can be based on a query that the master send
%, is assigned by the master based on the index $D$.
\begin{align}
I(D;Q_i,\tilde{A}_i,\boldsymbol{B})=0    
\end{align}
for any $i\in[N]$, where $\boldsymbol{A},\boldsymbol{B},D$ are sampled uniformly at random. Furthermore, we say the computing scheme is private and secure, if we also have 
\begin{align}
I(\tilde{A}_i;\boldsymbol{A})=0    
\end{align}
for any $i\in[N]$ when $\boldsymbol{A}$ is sampled uniformly random. 

Finally, for fully private batch matrix multiplication, the goal is to batch multiply $L$ pairs of matrices given two lists of inputs $\boldsymbol{A}=\{A^{(i,j)}\}_{i\in[L],j\in[M]}$ and $\boldsymbol{B}=\{B^{(i,j)}\}_{i\in[L],j\in[M]}$. The master aims to compute $\{{A^{(i,D)}}^\intercal B^{(i,D)}\}_{i\in[L]}$ given an index $D$, while keeping $D$ private. The rest of the computation follows similarly to the fully private and the batch distributed matrix multiplication frameworks.  
%The master send a query $Q_i$ to each worker $i$ based on the demand $D$. Then worker $i$ encodes %encoding functions for We require the requests to be private in the sense that 
Explicitly, we require that 
\begin{align}
I(D; Q_i,\boldsymbol{A},\boldsymbol{B})=0    
\end{align}
for any $i\in[N]$, where $\boldsymbol{A},\boldsymbol{B},{D}$ are sampled uniformly at random and $Q_i$ denotes the query the master send to worker $i$.

The achievability schemes for all these settings can be built based on coding ideas we presented earlier in this paper. In particular, by first pre-encoding each of the input matrices using any upper bound construction of $R(p,m,n)$, the task of batch-multiplying $L$ matrices is reduced to computing element-wise product of two vectors of lengths at most $LR(p,m,n)$. Then observe that in the second parts of all coding schemes we presented in earlier sections for non-batch matrix multiplication, we essentially provided linear codes that compute element-wise products of vectors of any sizes. By directly applying those proposed designs to the extended pre-coded vectors for batch multiplication, we obtain the needed computing schemes for proving Theorem \ref{thm:batch} where the achieved recovery threshold upper bounds are stated by swapping $R(p,m,n)$ into $LR(p,m,m)$.

%Solves open problems in subsequent works

%TODO: harmonic coding is better

%computational systematic

%although matdot can be viewed as a direct adaptation of coded convolution scheme in xx, [] is the first work to formally present in this context.

% \appendices

\section{Conclusion and Future Directions}

In this paper, we showed that entangled polynomial codes, as an effective approach for computing block matrix multiplication, can be applied beyond straggler mitigation. We investigated three important settings: secure, private, and batch distributed matrix multiplication, and demonstrated the effectiveness of entangled polynomial codes in providing unified solutions with orderwise improvements upon the state of the arts.
To demonstrate the coding gain, we focus on generalizing the second version of the entangled polynomial code, the one that achieves a recovery threshold of $2R(p,m,n)-1$. Note that similar to %entangled polynomial codes for
straggler mitigation, %which achieves optimal ``cubic" recovery thresholds $pmn+p-1$ when any of $p$, $m$, $n$ is small, 
where a ``cubic" recovery thresholds $pmn+p-1$ can be achieved when any of $p$, $m$, $n$ is small,
one should expect that similar optimal coding designs can also be developed for the  settings we studied in this work, and it is an interesting follow-up direction to establish those constructions. 

Another interesting direction is to consider general computation tasks beyond matrix multiplication. Entangled polynomial codes enables exploiting the idea that any bilinear function can be characterized by a rank-$3$ tensor, and any decomposition of the tensor reduces the bilinear function into batch evaluations of simpler computation tasks that can be assigned to distributed worker and effectively computed using Lagrange encoding. This approach directly generalizes to any multilinear functions. However, it remains open to design optimal codes for general computation with encoding and decoding functions satisfying complexity constraints.

\section*{Acknowledgements}
This material is based upon work supported by Defense Advanced Research Projects Agency (DARPA) under Contract No. HR001117C0053, ARO award W911NF1810400, NSF grants CCF-1703575, ONR Award No. N00014-16-1-2189, and CCF-1763673. The views, opinions, and/or findings expressed are those of the author(s) and should not be interpreted as representing the official views or policies of the Department of Defense or the U.S. Government. Qian Yu is supported by the Google PhD Fellowship.

%while requiring computing results from a factor of $2$  

%\section{Optimal linear codes for computing multilinear functions with distinct securit requirements per entries }\label{app:ss_mul}

	\bibliographystyle{ieeetr}
\bibliography{references}
\end{document}